\documentclass[12pt]{article}
\usepackage[english]{babel}
\usepackage{amsmath}

\newcommand{\rmd}{{\rm d }}
\newcommand{\rme}{{\rm e }}
\newcommand{\rmi}{{\rm i }}
\newcommand{\bra}{\langle}
\newcommand{\ket}{\rangle}

\title{On the Sharpness of Localization \\
of Individual Events in Space and Time} 
\author{Rudolf Haag \\[2mm] 
II.~Institut f\"ur Theoretische Physik, Universit\"at Hamburg \\
home address: D--83727 Schliersee--Neuhaus, Waldschmidtstr.~4b \\
e--mail: haag.rudolf@t-online.de}

\date{25 March 2013}

\begin{document}
\maketitle 

\begin{abstract} 
\noindent The concept of event provides the essential bridge from the realm  
of virtuality of the quantum state to real phenomena in space and
time. We ask how much we can gather from existing theory about their 
localization and point out that decoherence and coarse graining
-- though important -- do not suffice for a consistent interpretation 
without the additional principle of random realization.
\end{abstract} 

{\bf keywords:} Quantum events. Localization. Reality.

\newpage 

\section{Introduction}
What is an individual event, and what can we say about its space-time 
attributes? Viewed from orthodox quantum mechanics such questions do 
not have a straight forward answer. Niels Bohr criticizes the naive 
probability interpretation of the wave function, pointing out that we 
can assign no conventional attributes (like position and momentum) to 
an ``atomic object''. This tenet of Bohr's is central to his philosophy
of complementarity, \cite{BS,B}. It has not easily been accepted. Some 
ingenious constructions were presented in the attempt to save the idea 
that a particle always has some position 
(though unknown to us).\footnote{Best known is the suggestion 
by David Bohm [BD]. I would
also like to mention the work of Fritz Bopp in the years between 
1949 and 1955 \cite{BO}} We believe, however, that by now there is 
overwhelming evidence
for the truth of Bohr's statement. We must add to it  the claim that we 
can measure the position of a particle at given time  if we choose to do so.
Taken together this implies that
while such a ``measured value'' is not an attribute of the particle
itself, it is created by the interaction process of the particle with 
some other agent, which constitutes an essential part of the measuring 
device. It is an attribute of an ``event''.

	An individual event has an approximate location in
space--time. We would like to assign to it an intrinsic extension, the 
sharpness of its localization. We expect that in accordance with the 
uncertainty relations this sharpness increases with increasing 
energy--momentum transfer between the partners. Since macroscopic 
amplification is necessary for a ``measuring result'' to become
documentable  and thereby unquestionably real, we do not directly 
observe the sharpness of an individual event, but are limited by the 
resolution of the magnification (e.g.~a grain in a photographic
emulsion). Still, we claim that we can observe and register individual 
microscopic events and gather at least statistical information about 
their intrinsic extension.

	In order to avoid getting lost in vague generalities we focus 
first on a simple history: an $\alpha$--particle 
originating from the decay of 
an uranium nucleus passing through matter produces a succession of 
ionization events, documentable for instance by a track in a cloud
chamber. This stays within the regime of non-relativistic quantum
mechanics. The primary energy of the $\alpha$--particle of a few MeV
corresponds to a velocity $v$ of $10^9$cm/s and a deBroglie wavelength of 
$10^{-12}$cm; the energy transfer between the $\alpha$--particle and the ionized 
molecule is of the order of a few eV. Comparing the mean lifetime of 
the U${}_{238}$--nucleus (billions of years) with the detection time of 
the $\alpha$--particle by a scintillation counter (a few nano seconds) we 
see an enormous gap for the localization in time of single
events. There is the question about the coherence length of the
$\alpha$--particle prior the entrance into the cloud chamber and of the 
relevance of this length for the duration of the ionization process. 
In the standard approach one starts with a plane wave (infinite 
coherence length) representing a constant current density and 
yielding a number of events per unit time, without being concerned 
about the duration of the individual event. 

	The questions raised here and illustrated in this simple 
example appear to be of fundamental importance in various respects. 
First of all, there is the task of defining a physical counterpart to 
the abstract notion of a point in space--time. We expect that in
agreement with Bohr's tenets and Wigner's objections mentioned in the 
next section this is not obtainable as a position of a particle but 
can be approximated by a high--energy event. Thus the concept of
``event'' provides the necessary bridge to the finer features of space--time. 

	Secondly, there is the question of what the probabilities 
in quantum physics really refer to. Probabilities for measuring
results? Or probabilities for events? 

	Third, there is the reality question. It has a long history, 
beginning with the discussions between Bohr and Einstein, illustrated
in the paradoxon of Schr\"odinger's cat, and accentuated by the 
experiments of entanglement and violation of Bell's inequality. As a 
consequence, doubts were raised about the existence of ``mind
independent'' reality in physics with the assertion that either 
reality or locality has to be sacrificed. The view proposed here 
suggests that this challenge concerns the concept of an individual 
particle. It is neither real nor localized, whereas both reality 
and locality can be attributed to an individual event. This 
discussion leads us to a critical assessment of the notion of
``observable'' and of the transition from possibilities to individual facts.

\section{The track of an $\alpha$--particle}

In  his book ``Die physikalischen Prinzipien der Quantentheorie'' 
(1930) \cite{He} which appeared soon after the creation of the 
Copenhagen interpretation of the formalism of quantum mechanics,
Heisenberg devoted one section to the discussion of the track of 
an $\alpha$--particle in a Wilson cloud chamber.

Heisenberg's main objective was to show that the track is 
approximately a straight line in the direction of the incoming
momentum. 
For this purpose he restricted attention to two water molecules at 
fixed positions $\mathbf{X}_i$ ($i$=1,2) and calculated the ionization 
process in  perturbation theory, representing the incoming 
$\alpha$--particle by a plane wave. It is noteworthy that he 
approached the issue from two distinct points of view.
First he considered the individual water molecules as 
instruments for a position measurement of the $\alpha$--particle. 
In this case one assumes that the ionization of an individual 
water molecule is always followed by the formation of a droplet. 
The size of the droplet limits the attainable precision of the 
position measurement.
Heisenberg is not concerned with the step from the ionization of 
a single molecule to the appearance of a droplet. He takes it for 
granted that it can be regarded as a one--to--one connection. This is 
probably well justified. It is a typical feature shared with many 
detection processes. We can usually identify some micro--event responsible for
triggering an avalanche effect producing the amplification needed for
perception by our senses. This feature is responsible for our ability 
to register individual microscopic processes. Still a thorough 
discussion of this relation between a microscopic event and its 
amplification under realistic circumstances would be highly desirable.

 In the second point of view he considers the $\alpha$--particle 
together with two water molecules as the “physical system” and 
computes the time development of the joint wave function.

The phenomenon offers an illustration of a number of basic 
questions concerning the understanding of quantum theory. 
Among them: \\[2mm]
(i) \ Comparison between the two points of view and the notion of 
consistent histories  \\[2mm]
(ii) \ Localization of individual events. 
Here we must distinguish two parts: \\[1mm] 
(a) \ the effective range of the interaction.
It depends on the momentum transfer between the partners, limits the 
starting distance between the outgoing particles and is related to the 
cross section.
If one of the partners is assumed to be fixed in space 
(as in Heisenbergs scenario) it gives the transverse localization. \\[1mm] 
(b) \ Localization of the collision center in space and time. \\
It is restricted by the overlap of the wave functions of incoming 
particles. These depend, first of all, on the effective coherence 
lengths which are the inverses of the momentum uncertainties. 
Here it is important to realize, that this includes the uncertainty 
in the control of the momentum, which will usually be some small 
fraction of the total energy--momentum. In the high energy regime 
this alone may suffice to bring the effective coherence length below
the limits of experimental resolution. See Subsection \ref{A.3} and appendix.

The increase of sharpness of localization with increasing energy 
transfer becomes dramatic in high energy processes. There we may 
expect that the localization of the vertex is much sharper than 
the limit of experimental resolution. We cannot attempt to address 
this problem here but just illustrate its fundamental significance 
by an episode which casts a spotlight on a controversial issue: 
Some decades ago during a talk on local observables  with the 
assumption of strict locality (Einstein causality). E.P.~Wigner 
who was in the audience became increasingly restless and ultimately 
objected: 

"We have shown many years ago that in relativistic 
quantum theory the measurement of positions does not have a 
precise meaning". 

In a subsequent private discussion the 
following exchange took place. 

\vspace*{2mm} 
\begin{tabular}{lp{11cm}}
R.H.: & "Did you refer to the paper by Newton and Wigner?" \\
E.P.W.: & "Oh, you have read this paper!" \\
R.H.: & "Yes, but why do you want to mark a point in space by the position 
of a particle?" \\
E.P.W.: & "How else do you want to do it ?" \\
\end{tabular} 

\vspace*{2mm} 
\noindent  How else indeed!
The answer appears to be that the operational approach to a 
point (in fact a point  in space-time) can only be realized by a 
high energy interaction process involving several particles, an
event. 
The notion of “events” must be regarded as an independent primary 
concept intimately tied to relations in space-time.

\subsection*{A.1 \ Comparison between the two points of view} 
It is part of the orthodox Copenhagen interpretation that 
we must introduce a cut between the physical system under 
consideration and the measuring instruments used by the observer. 
There is some arbitrariness in the choice of the cut. In the above 
example the two points of view discussed by Heisenberg correspond 
to different choices of the cut. We may ask about the legitimacy 
of the placement of the cut. Considering the molecules as measuring
instruments one assumes that the first ionization is a completed
process 
before the second begins and its result is adequately described by a 
projector $P_1$ associated to a small volume around the center of 
the molecule. A test for the legitimacy of the assumed placement 
of the cut has been proposed by Griffiths \cite{Gr} with the notion 
of "consistent histories". The assumed "history" is the occurrence 
of two subsequent events (associated with the projectors $P_1$ at 
time $t_1$ and $P_2$ at time
$t_2$). Writing $P_1'=1-P_1$ for the complement (negation) of $P_1$ 
the statistical matrix $\rho $ may be written as 
$$\rho =P_1\rho P_1 +P_1 \rho P_1' + P_1' \rho P_1 +P_1' \rho P_1'.
$$
The measurement of $P_1$ cancels the off-diagonal parts and 
leads to the new statistical matrix $$\rho '=P_1 \rho P_1 +P_1' \rho  P_1'.$$
If the probability for a subsequent measurement $P_2$ is not 
affected by the preceding performance of $P_1$ we must have 
$$Tr P_2\rho P_2 =Tr P_2 \rho ' P_2$$
which can be simplified to the "consistency condition"
$$ Tr P_2 \, [P_1, [\rho, P_1]]=0 \, .$$
Of course consistency of an assumption does not yet guarantee its
truth. 
Somewhat more satisfactory is the direct comparison of the joint 
probability for the two events obtained in the two points of view. 
This amounts to the question as to whether the joint probability 
(calculated by the second method) can be broken up into a product 
of the probability for the first event times the conditional 
probability for the second subject to the occurrence of the first. 
Some remarks about this will be given at the end of A3. 
In any case it is clear that the separation of the phenomenon 
into a succession of events  needs that the distance of the 
two ionized molecules (mean free path of the $\alpha $-particle) 
is sufficiently large. 

\subsection*{A.2 \  The transverse localization in space \label{A.2}} 

In Heisenberg's scenario the localization of an event in space is due to the 
locality of the interaction Hamiltonian. Heisenberg estimates 
the diameter of the scattered wave of the $\alpha $-particle 
after its first encounter with a water molecule to be of 
molecular dimension. He does not specify the interaction he used. 
The relevant interaction for the process in lowest order is the 
Coulomb force between the $\alpha $-particle and the electron to 
be ejected. One may wonder how such a long range force succeeds 
in confining the event to a region of molecular size starting 
from a plane wave corresponding to initial complete ignorance. 
The jump is a consequence of the demanded energy momentum transfer. 
This can be seen from the following very rough, classical, 
order of magnitude estimate. 

Suppose the trajectory of the $\alpha$-particle passes at a 
distance $d$ from the center of the molecule. If  $d$ is somewhat 
larger $R$, 
where $R$ is the radius of the molecule, the force exerted by the 
$\alpha$-particle on the electron is smaller than $\frac{2e^2}{d^2}$ 
throughout the process. Under the influence of this force the 
electron must suffer a displacement of at least $R$ and the energy 
transfer to it must exceed the binding energy 
$E_0$, so $\frac{ 2 e^2 R}{d^2} > E_0$.
For $E_0 = 2eV$, using $e^{2} \sim 2 \cdot 10^{-7} cm \cdot eV $ 
we get $\frac{d^{2}}{R}<4 \cdot 10^{-7} cm$. Otherwise no ionization is 
possible and this effective range gives the transverse spatial localization 
of the event. In the quantum-mechanical calculation the decrease 
of the effective range with increasing energy transfer may be 
traced to the damping effect of increasingly rapid 
oscillations in the integrals for the matrix elements.

\subsection*{A.3 \   Localization in time \label{A.3}} 
The longitudinal and temporal localizations of an event are 
less clearly visible than the transversal one. The classical 
picture sketched above indicates that the longitudinal range 
$d_\mathrm{long}$ will be a few times larger than the 
transversal one, $d_\mathrm{tr}$, but remains in the same order of magnitude. 
The localization in time is then $d_\mathrm{long}/v$ unless we 
are dealing with a resonance phenomenon leading to a long delay 
time (which is not the case in our example).

In quantum theory there arises the question about the coherence 
length $l_\mathrm{c}$ of the wave packet describing the 
$\alpha$--particle prior to its entrance into the Wilson chamber. 
Presumably the dominant restriction is due to the thermal noise 
to which the uranium probe is subjected. This produces some 
random oscillatory motion of the uranium atoms with some mean 
velocity $ v_{u} $ which via Doppler effect leads to an 
uncertainty of the frequency of the matter wave: \
$ \Delta \nu_{\alpha}=\frac{v_{U}}{\lambda} $ ,\ 
where $\lambda$ is the wavelength of the matter wave. This 
formula can also be simply understood in the particle picture, 
where the energy uncertainty is \ 
$\Delta E_\mathrm{\alpha}=\Delta 
(\frac{mv_\mathrm{\alpha}^{2}}{2})=p_\mathrm{\alpha}v_\mathrm{U}$ .\ 
We estimate $v_{U} \sim 10^4 \frac{cm}{s}$ for the probe at room 
temperature. This corresponds to a relative uncertainty of the 
momentum $\frac{\Delta p}{p} \sim 10^{-5} $ and a coherence 
length $ l_{c} \sim 2 \cdot 10^{-7}\, cm$.

Irrespective of the value of the effective coherence length 
we want to show that if it is much larger than $d_{long} $ 
it has little influence on the time duration of an 
individual ionization process.
We shall calculate this in Heisenbergs scenario in perturbation theory  
evaluating the time dependent approach explicitly and for finite time intervals.
This computation is elementary and standard. 
It is only presented to illustrate some of the points mentioned. 
It will allow the comparison between the classical and the 
quantum  prognosis for $d_\mathrm{tr}$ and $d_\mathrm{long}$.

In order to keep the effort within reasonable bounds and to 
focus on the essentials we replace the water molecule by an 
alkali atom, where we have a clear distinction of the electron to be
ejected. \\[2mm]
\textbf{Notation:}

\begin{itemize}
\item[-] We chose units so that  $\hbar=1$ 
\item[-] State vectors are denoted by bold face Greek capitals 
like $\mathbf{\Psi}, \mathbf{\Phi}$ or by Dirac kets like 
$|\mathbf{p} \, \mathbf{q}\rangle$; 
wave functions by ordinary Greek letters like $\psi, \phi$; 
\item[-] Position resp.~momentum of the $\alpha$-particle: 
$\mathbf{x} = ( x_1, x_2, x_3 ),\ \mathbf{p}$; \\
  Position of the center of the atoms: $\mathbf{X}_i \ (i=1,2)$; \\ 
  Positions resp.~momenta of electrons: $\mathbf{X}_i+\mathbf{\xi}_i,\ 
    \mathbf{q}_i$; 
\item[-] Quantities in the interaction picture are adorned with a hat: 
$\mathbf{\hat{\Psi}}(t), \hat{\psi} (t)$; \ 
in the Schr\"odinger picture by a suffix $\mathrm{S}$: 
    $\mathbf{\Psi}_\mathrm{S}$; 
  the two pictures coincide for $t=0$; 
\item[-]  
$\mathbf{\hat{\Psi}}(t)=\rme^{\rmi \mathrm{H}_0 t} \mathbf{\Psi}_\mathrm{S}(t)$;
\ The Hamiltonian 
$\mathrm{H}_0$ includes the Coulomb energy between electron and atomic
core; \
The interaction Hamiltonian consists of the Coulomb energy 
between $\alpha$ and electron  
$\mathrm{V}_i=\frac{2e^2}{|\mathbf{x}-\mathbf{X}_i-\mathbf{\xi}_i|}$, 
as well as the Coulomb energy of the $\alpha$-particle and the atomic
cores. 
The latter will however give no contribution to the matrix elements 
considered here and thus will be dropped in the sequel. 
\item[-]  The basis we use consists of \\ 
$|\mathbf{p}\rangle 
\otimes \mathbf{\Phi}_{10} \mathbf{\Phi}_{20}= |\mathbf{p}  \, 
\mathbf{\Phi}_{10} \mathbf{\Phi}_{20} \rangle; \ |\mathbf{p} \, \mathbf{q} 
\rangle \otimes \mathbf{\Phi}_{1+} \mathbf{\Phi}_{20}; \ |\mathbf{p} \,
\mathbf{q}_1 \mathbf{q}_2\rangle \otimes \mathbf{\Phi}_{1+} 
\mathbf{\Phi}_{2+}$ \\ 
all referring to t=0; \ 
Here $\mathbf{\Phi}_i$ is the state of the atom which may be 
either in the neutral state $\mathbf{\Phi}_{i0}$ or ionized as 
$\mathbf{\Phi}_{i+}$ accompanied by a free electron
$|\mathbf{q}\rangle$. \
This indicates the 
temporal sequence of processes: atom $i=1$ 
is the one ionized first (in first order of perturbation expansion); \\ 
Since $\mathrm{H}_0$ includes the Coulomb energy between electron 
and core, $|\mathbf{q}\rangle$ is an eigenfunction in the 
continuous spectrum with asymptotic momentum~$\mathbf{q}$;
\end{itemize}

\noindent {}From   the equation of motion 
\begin{equation}
	\rmi \frac{\rmd \mathbf{\hat{\Psi}}(t)}{\rmd t} = 
\hat{\mathrm{V}}(t) \mathbf{\hat{\Psi}} (t),
\end{equation}
where 
$\mathbf{\hat{\Psi}} (t) = \rme^{\rmi \mathrm{H}_0 t}
\mathbf{\Psi}_\mathrm{S}(t); \  
\mathrm{V}(t)= \rme^{\rmi \mathrm{H}_0 t} \mathrm{V} \rme^{-\rmi \mathrm{H}_0 t}$
we get in first order 
\begin{equation}
\hat{\psi}_{+}^{(1)} (\mathbf{p}_1,\mathbf{q}_1,t_1)
=\int\limits_{0}^{t_1} \rmd t \int \rmd^3 \mathbf{p}\ M(\mathbf{p}-
\mathbf{p}_1,\mathbf{q}_1) \psi_0 (\mathbf{p}) \rme^{\rmi (E_1-E)t},
\label{integration}
\end{equation}
with
\begin{equation}
M(\mathbf{p}-\mathbf{p}_1,\mathbf{q}_1)=\int\limits \rmd^3 
\mathbf{k} \, \rmd^3 \mathbf{x} \frac{2 e^2}{\mathbf{k}^2} \rme^{\rmi  
(\mathbf{k+p-p_1})(\mathbf{x}-\mathbf{X}_1)} \langle \mathbf{q}_1| 
\rme^{\rmi \mathbf{k} \mathbf{\xi}_1} | \varphi_0 \rangle,
\end{equation}
\begin{equation}
E_1 = \frac{\mathbf{p}_1^2}{2 m_\alpha} + \frac{\mathbf{q}_1^2}{2
  m_\rme}\ 
\text{and}\ 
E = \frac{\mathbf{p}^2}{2 m_\alpha} - | E_0 |.
\end{equation}
$| \varphi_0 \rangle$ is the ground state of the electron.
Note that we could introduce
\begin{equation}
\widetilde{M}(\mathbf{x}-\mathbf{X}_1,\mathbf{q}_1)=
\int\limits_{|k_i| > \mathbf{\Delta}_i} \rmd^3 
\mathbf{k} \frac{2 e^2}{\mathbf{k}^2} \rme^{\rmi  
\mathbf{k}(\mathbf{x}-\mathbf{X}_1)} \langle \mathbf{q}_1| 
\rme^{\rmi \mathbf{k} \mathbf{\xi}_1} | \varphi_0 \rangle,
\end{equation}
where $\mathbf{\Delta}$ is the demanded minimal momentum transfer. 
This introduces the gap in the $\mathbf{k}$-integration, which is 
responsible for the short extension of $\widetilde{M}$ as a 
function of $\mathbf{x-X}$ of order 
\begin{equation}
\mathbf{d}_i \approx \frac{\pi}{\mathbf{\Delta}_i}.
\end{equation}
This may be compared to the classical estimate above in 
Subsection A.2.

To get an intuitive feeling it is good to look at $\psi_{+}^{(1)}$ in 
Schr\"odinger's position representation:
\begin{align}
\psi_{+\mathrm{S}}(\mathbf{x_1,\xi_1}, t_1) &= \int\limits \rmd 
\mathbf{p}_1 \rmd \mathbf{q}_1 \rme^{-\rmi E_1 t_1}
\psi_{+}(\mathbf{p_1,q_1},t_1) 
\rme^{i \mathbf{p_1 x_1} + i \mathbf{q_1 \xi_1}} \\
	&= \int \rmd \mathbf{p} \int \rmd \mathbf{p}_1 \rmd
        \mathbf{q}_1 
\rmd t M(\mathbf{p-p_1,q_1}) \psi_0 (\mathbf{p}) \rme^{\rmi \chi}.
\end{align}
The phase
\begin{equation}
\chi = \mathbf{p_1 x_1 + q_1 \xi_1 - E_1 (t_1-t)-Et}
\end{equation}
is a quickly changing function of $\mathbf{p_1, q_1}, t$ and we 
shall first evaluate it in the stationary phase approximation. 
$\chi$ has a stationary point for $\mathbf{p_1 = \overline{p}_1,\ q_1
  = 
\overline{q}_1}$ and $t=\overline{t}$ with
\begin{equation}
\overline{\mathbf{p}}_1 = \frac{m_\alpha 
( \mathbf{x_1 - X_1})}{t_1-\overline{t}},\  
\overline{\mathbf{q}}_1 = \frac{m \mathbf{\xi}_1}{t_1-\overline{t}},
\end{equation}
\begin{equation}
t_1 - \overline{t} = \sqrt{\frac{A}{2 E}}, \label{eq:t_stat}
\end{equation}
where 
\begin{equation}
A = m_\alpha^2( \mathbf{x_1-X_1})^2 + m_\rme \mathbf{\xi}_1^2.
\end{equation}
We get
\begin{equation}
\psi_{+\mathrm{S}}(\mathbf{x_1,\xi_1}, t_1) = C \int \rmd \mathbf{p} 
M(\mathbf{p-\overline{p}_1,\overline{q}_1}) \psi_0(\mathbf{p}) \rme^{i \chi_1},
\end{equation}
where
\begin{equation}
\chi_1 = \sqrt{2EA} - E_1 t_1.
\end{equation}

Noting that the  transverse motion $\mathbf{x}_\perp$ is negligible 
compared to the longitudinal one, that the contribution of 
$\mathbf{\xi}_1$ to A may be ignored due to the mass ratio 
$\frac{m_\rme}{m_\alpha}$ and finally, that the binding energy 
is negligible compared to the kinetic energy of the $\alpha$-particle, we have
\begin{equation}
\chi_1 \approx p_\mathrm{z} (z_1 - Z_1) - \frac{p^2}{2 m_\alpha} t_1.
\end{equation}
\\
\textbf{Comments}
\\
The selection of $\bar t$ (\ref{eq:t_stat}) is just energy conservation $E_1 = E$.
\\
The quality of the stationary phase approximation can  be 
estimated by the widths $\Delta_p = 
\sqrt{\frac{m_\alpha}{t_1-\overline{t}}}$, $\Delta_q = 
\sqrt{\frac{m_\rme}{t_1-\overline{t}}}$ and $\Delta_t = 
({\frac{A}{2 E^3}})^{1/4}$ following from the second derivatives of $\chi$.
\begin{equation}
\Delta_t = \! \Big( \frac{\partial^2 \chi}{\partial t ^2} \Big)^{-1/2}
\! \! = \Big( \frac{2E^3}{A} \Big)^{-1/4} 
\! \! =\Big( \frac{z-Z}{E_\alpha v_\alpha} \Big)^{1/2}
\! \!  = \Big( \frac{10^{-6} cm}{10^{21} s^{-1} 10^9 cm s^{-1} } \Big)^{1/2}
\! \! = 10^{-18} s.
\end{equation}
The comparison to the passage time 
$\frac{10^{-7} cm}{10^9 cm/s} = 10^{-16} s$ shows that for the 
$t$-integration the stationary phase method is good. However, 
for the $\mathbf{p}$-integration this method is not so good. 
We shall not use it for further calculations and only use it 
as an indication of the qualitative behavior.

To estimate the influence of the coherence length $l_\mathrm{c}$ we may put
\begin {equation}
\psi_0(\mathbf{p}) = \delta (\mathbf{p}_\perp) \rme^{-(p-\overline{p})^2 l_\mathrm{c}^2}.
\label{factor}
\end{equation}
For $l_\mathrm{c} < d_\mathrm{long}$ the 
factor (\ref{factor}) cuts part of 
the integration of (\ref{integration}) and thus affects the final momentum 
distribution, so that it approaches the semi-classical model in 
which a classical point charge moves past the atom. For
$l_\mathrm{c}>5 
d_\mathrm{long}$ this effect becomes negligible and the momentum 
distribution has reached M. When $\psi_0$ approaches a plane wave 
($l_\mathrm{c} \rightarrow \infty$) the momentum distribution does 
not change any more but the pattern of the scattered wave becomes 
stationary describing a constant flow. This does not mean, however, 
that the individual process lasts infinitely long. The time 
$\overline{t}$ at which the filament, which is responsible for 
the amplitude of the scattered wave at position $X_1$ at time $t_1$, 
interacts with the atom is
\begin{equation}
	\overline{t} = t_1 - \sqrt{\frac{A}{2E}} \approx t_1 - \frac{z_1}{v}.
\end{equation}

We now reached the central point appearing in any discussion of the 
measuring process. Namely  the transformation of statistical 
information in an ensemble, encoded in the quantum state 
$\hat{\mathbf{\Psi}}$, 
to probabilities for the occurrence of individual events (facts).

What does our calculation of the deterministic propagation of matter 
waves tell us about the phenomenon under consideration? In this we 
must at some stage perform the jump from amplitudes to probabilities. 
This means the squaring of the wave function in some basis. 
The choice of when to do that reflects the judgment as to when we 
may consider an individual process to be completed. The choice of 
the appropriate basis corresponds to the answer to the 
question "Probability for what?" among many complementary 
possible choices. This is the general problem whose discussion 
we defer to the next section. For the circumstances considered 
here the practical answer follows by common sense.

If we divide $l_\mathrm{c}$ into pieces $l_\mathrm{c}^{(i)}$ of 
length say $5 d_\mathrm {long}$ then each of these filaments 
$\hat{\mathbf{\Psi}}_{+(i)}$ passes the atom at a different mean time 
$\overline{t}^{(i)}$. After passing, it leaves it with a momentum 
distribution M, which remains unchanged in the remaining interval 
till $t_1$. The position of its center has accordingly shifted to the 
neighborhood of $z_1 \approx v (t_1 - \overline{t}^{(i)})$. We then
have 
at time $t_1$ a sum of wave packets $\mathbf{\Psi}_{+(i)}$ centered at 
different places. It is clear that in the considered setting 
(the Wilson chamber) no interference between the different wavelets 
$\mathbf{\Psi}_{+(i)}$ is possible. Thus the initially assumed pure state 
$\mathbf{\Psi}_0$ is effectively transformed into the mixture 
$\sum\limits_i | \mathbf{\Psi}_{+(i)} \rangle \langle \mathbf{\Psi}_{+(i)}|$ and 
$|| \mathbf{\Psi}_{+(i)} ||^2$ may be regarded as the probability of an 
individual ionization process occurring at time $\overline{t}^{(i)} 
\pm \tau$ with $\tau \sim 5 \frac{d_\mathrm{long}}{v}$.

This conclusion agrees with the standard practice of calculating the 
cross section. There, one starts with the part of the S-matrix element 
in the momentum representation, which is essentially $M
\delta(E-E_1)$, 
going over to a probability by straightforward squaring. The fact that 
$\delta^2 = \infty$ reminds us that we have to change the
normalization 
from the passage of a single $\alpha$-particle to an $\alpha$-particle 
beam with finite current density J.
\begin{equation}
W=\int \rmd E |M|^2 \delta (E-E_1)
\end{equation}
is then the probability of ionization per unit time related to the 
cross section by
\begin{equation}
W = \sigma J.
\end{equation}
This gives the same information about the frequency of an event as 
in our more elaborate computation above with the only difference that 
it does not contain any information about the duration of the
individual 
process.

It remains to discuss whether one can regard the track as a history 
of separate events or whether one must consider it as a single 
complex event. We therefore study whether the probability 
for a composite process of two successive ionizations can be broken 
up into a product of a probability for the first event and a
conditional 
probability for the second, subject to the occurrence of the first.
The probability amplitude for the composite process is in lowest order given by 
\begin{equation}
\hat{\mathbf{\Psi}}_{++}^{(2)} = \int\limits_0^{t_2} \rmd t_1 \hat{V}(t_1) 
\int\limits_{0}^{t_1} \rmd t \hat{V}(t) \mathbf{\Psi}_0 = \int\limits_0^{t_2} 
\rmd t_1 \hat{V}(t_1) \hat{\mathbf{\Psi}}_{+}^{(1)} (t_1)
\end{equation}
or
\begin{equation}
\hat{\psi}_{++}^{(2)} = \int\limits_0^{t_2} \rmd t_1 \rmd \mathbf{p}_1 
M(\mathbf{p_2-p_1,q_2}) \rme^{\rmi(E_2-E_1)t_1} \rme^{-\rmi 
(\mathbf{p_2 - p_1})\mathbf{X}_2} \hat{\psi}_{+}^{(1)}(\mathbf{p_1,q_1},t_1).
\label{eq:psi2_int}
\end{equation}

In order to cut this into a probability for the first 
ionization process times a transition probability, one 
has to replace $\hat{\psi}^{(1)}_+ (t_1)$ in the integral 
(\ref{eq:psi2_int}) by its asymptotic value, which is 
reached in the time interval $\tau$. This approximation 
is good if $\tau \ll t_1$ or $L \gg 5 d_\mathrm{long}$, where $L$ is 
the mean free path between the two ionization processes. Then
\begin{equation}
\hat{\psi}^{(2)} = \int\limits \rmd \mathbf{p}_1
M(\mathbf{p_2-p_1,q_2}) 
\delta(E_2 - E_1) \rme^{-\rmi (\mathbf{p_2 - p_1})\mathbf{X}_2} 
\frac{ \hat{\psi}^{(1)}_\mathrm{as} }{||\hat{\psi}^{(1)}_\mathrm{as}||} 
\end{equation}
is interpreted as the conditional probability for the ionization for 
the atom at position $ \mathbf{X}_2$ due to the normalized incident 
wave $\frac{ \hat{\psi}^{(1)}_\mathrm{as}
}{||\hat{\psi}^{(1)}_\mathrm{as}||} $ 
emanating from the first atom.

\vspace*{2mm} 
\noindent \textbf{B \ 
Many atoms with unknown positions, many $\alpha$--particles} \\
The realistic situation is that the initial state of the cloud 
chamber consists of many atoms with unknown positions of their 
centers of mass. We may introduce creation operators $a^*(\mathbf{X})$ 
for an atomic core with center $\mathbf{X}$, $b^*(\mathbf{\xi})$ for 
an electron at position $\xi$ and $c^*(\mathbf{x)}$ for an 
$\alpha$-particle at $\mathbf{x}$. 
We distinguish wave functions for atomic cores $A(\mathbf{X})$ 
by indices $i, j, ...$ using the same index for the wave function 
of the associated outer electron and indices $\rho, \sigma, ...$ 
for wave functions of the $\alpha$-particles. For the initial state we write 
\begin{equation}
\mathbf{\Psi}^{(0)}= \prod\limits_{i, \rho} a^*_i b^*_i c^*_\rho | 0 \rangle,
\end{equation}
where $ | 0 \rangle $ is the vacuum state and
\begin{align}
 a^*_i b^*_i &= \int A_i (\mathbf{X}) a^*(\mathbf{X}) 
b^*(\mathbf{X+\xi}) \phi_0 (\xi) \, \rmd \mathbf{\xi} \, \rmd \mathbf{X} \\
c^*_\rho &= \int c^*(\mathbf{x}) \Psi^{(0)}_\rho (\mathbf{x}) \, \rmd \mathbf{x}.
\end{align}
As the interaction we take 
\begin{equation}
V=\int \frac{e^2}{| \mathbf{x-\xi}|} c^*(\mathbf{x}) 
b^*(\mathbf{\xi}) c(\mathbf{x}) b(\mathbf{\xi}) \rmd \mathbf{x}\,\rmd 
\mathbf{\xi}\,,
\end{equation}
leaving aside again the interactions between $\alpha$-particles 
and atomic cores.
\\ The creation operators together with their adjoint 
annihilation operators satisfy the standard canonical 
commutation (resp.~anticommutation) relations.
The first order approximation gives
\begin{equation}
\hat{\mathbf{\Psi}}^{(1)} (t_1) = \int_0^{t_1} \hat{V}(t) \mathbf{\Psi}^{(0)}(t) \rmd t
\end{equation}
commuting the destruction operators in $\hat{V}$ to the right 
till they hit the vacuum, we obtain a sum of terms for 
$\hat{\mathbf{\Psi}}^{(1)}$ each of which corresponds to a 
pairing of an $\alpha$-particle with an atom.
\begin{equation}
\hat{\mathbf{\Psi}}^{(1)}(t_1) = \sum \hat{\mathbf{\Psi}}^{(1)}_{i,\rho},
\label{eq:sum_irho}
\end{equation}
with
\begin{alignat}{1} 
& \hat{\mathbf{\Psi}}^{(1)}_{i,\rho} \ = \\
& \rme^{\rmi H_0 t_1} \! 
\int\limits_0^{t_1} \! \rmd t \! \int \! \rmd \mathbf{X} \rmd \mathbf{\xi} 
\rmd \mathbf{x}\frac{2e^2}{|\mathbf{x-\xi}|} A_i (\mathbf{X},t) 
\phi_0(\mathbf{\xi-X})\rme^{-i E_0 t} \psi_\rho^{(0)}( \mathbf{x},t) 
a^*(\mathbf{X}) b^*(\mathbf{\xi}) c^*(\mathbf{x}) |0\rangle\, . \notag
\end{alignat}
We note first that the different terms 
$\hat{\mathbf{\Psi}}^{(1)}_{i,\rho}$ in the sum (\ref{eq:sum_irho}) 
are incoherent because their phase relation is in principle 
uncontrollable. Thus we should replace $\hat{\mathbf{\Psi}}$ 
by the statistical matrix
\begin{equation}
\sum\limits_{i,\rho} | \hat{\mathbf{\Psi}}^{(1)}_{i,\rho} \rangle 
\langle \hat{\mathbf{\Psi}}^{(1)}_{i,\rho} |,
\end{equation}
which means that each individual event realizes one of the 
possible pairings and that this happens with a probability 
\begin{equation}
W_{i, \rho} = ||\hat{\mathbf{\Psi}_{i,\rho}}||^2.
\end{equation}
The wave function of the state $\hat{\mathbf{\Psi}}_{i,\rho}^{(1)}$ 
differs  from the wave function $\psi_{+}^{(1)}$ in Subsection 
\ref{A.3} only because the atomic core is now not fixed at some
position, 
but described by the wave function $A_i (\mathbf{X})$ and therefore 
spread out and movable. This can be broken up into a sum of
effectively 
incoherent terms by a decomposition of $A_i$ into different 
parts of the phase space similar to the division of $l_\mathrm{c}$ 
into pieces $l_\mathrm{c}^{(i)}$ in Subsection A.3. We shall, 
however, not pursue this. 

In second order we obtain contributions ${\bf \Psi}^{(2)}_{i\rho,k\sigma}$ 
corresponding to the pairing $i\rho$ in the first event and$ \sigma$ 
in the second. If all indices are different there is no distinction of the 
temporary order. They correspond to the beginnings of different tracks. In 
this case there are two terms with the same indices differing by the 
permutation of $i$ and $k$ These are coherent (effect of Hanbury-Brown and 
Twiss). All others are incoherent. The composite event of two subsequent 
ionizations in one track corresponds to the case of equal indices 
$\rho=\sigma$.

\section{The reality issue}

\subsection*{A \  Experiment and Theory}

\noindent Wolfgang Paul liked to say that he was engaged in the real part of 
physics (experiments), whereas the additional ``i'' in the name of his 
colleague Wolfgang Pauli indicated the imaginary part of physics 
(the theory). This is more than a nice joke. Experimentalists have to 
regard their efforts as a dialogue with an outside world called
nature, and the individual observed phenomena as ``facts'', i.e. 
irreversible elements of reality, where ``nature'' and ``reality'' are 
essentially synonymous. This outside world is felt to be distinct from 
the human mind, obeying laws independent of our will. 

This corresponds to a dualistic picture of the universe, with two 
co-existing parts: human consciousness and will on one side, and 
nature on the other. The question about the relation of these 
two parts, known as the ``mind-body-problem'', has been a topic 
in philosophy for ages. Attempts at unification by eliminating 
one of the two sides led to the two extreme positions of 
idealism and materialism. 
	
For the purpose of physics it is not relevant to which ideology 
(if any) one adheres. The essential criterion for accepting an element 
of consciousness as the cognition of a counterpart in reality is the 
consensus between many observers, which lifts the impression from one 
individual consciousness to a collective one. If this is satisfied, 
the agreement of all people concerned is adequate for treating the 
said observation ``as if'' it were an element of an outside world, and 
there is no risk in omitting the ``as if'', but a great gain in the 
simplicity of the language. 

	There are, however, two points responsible for the deviation
of the description of a phenomenon in quantum physics from a fully
realistic account as one is used to give in everyday life and in 
classical physics. They are: \smallskip

\noindent (1) the impossibility of direct observation of microscopic objects or
 events and of reaching a consensus about the precise attributes of 
associated phenomena.

\noindent (2) the unavoidable and uncontrollable interaction of the observer 
with the observed object, making the subject-object separation 
fuzzy. \smallskip 

We should therefore look closely at the facts about which consensus 
can be reached and the way the difficulties are handled in experiment 
and theory. In this analysis we shall restrict attention to 
``Particle Physics'' i.e. the regime of extremely low density. 
There the simplest experimental set up consists of a source, emitting 
a beam of particles of known species, an area of manipulation  by 
external fields etc. and finally an array of detectors.

In our context more relevant are collision experiments in which there 
is in addition a reaction area with a target or, in a cross beam 
experiment, a second source producing another beam of particles.

A detector signal, being documentable, is real,
macroscopically localized and it is an irreversible fact. 
If we believe in an essentially deterministic connection between the 
signal and a microscopic ionization process as its cause we can extend 
reality to the microscopic event, though the attributes of it are 
hidden from our direct observation.

More subtle is the assessment of our description of the beam and the 
probability assignments for the events.
I want to discuss this very carefully at the expense of a somewhat 
tedious elaboration of detail and the recounting of generally known things.

	The interaction between particles within the beam is
negligible and the intensity may be reduced down to a flux of 
one particle per unit time. Therefore the beam may be regarded 
as a Gibbs ensemble of individual particles.

In quantum theory it is mathematically characterized by a positive 
operator with unit trace acting in Hilbert space. It is called the 
quantum state of the particles in the beam. The set of such operators 
is a convex set: for any pair of such operators $\rho_1$, $\rho_2$ and 
any positive number $\lambda < 1$ the convex 
combination $\lambda \rho_1 + (1 - \lambda) \rho_2$ belongs again to this
set. It is the mixture of $\rho_1$, $\rho_2$ with weights 
$\lambda$, $(1 - \lambda)$. This convex set 
has extreme points, namely one--dimensional projectors, which cannot be 
decomposed further.
Every state can be described as a convex combination of pure states:
\begin{equation}
\rho = \sum \lambda_k \, P_k \, , \quad  \sum \lambda_k = 1 \, .
\end{equation}
Each pure state appearing in such a decomposition characterizes a 
subensemble in the beam. It is tempting to believe that the pure state
$P$ describes a property of all particles belonging to the respective 
subensemble. This picture is, however, not advisable. 
The decomposition of the original mixed state into pure components 
is highly non unique. A striking example is afforded by the
equilibrium state of an ideal gas in a large vessel. It is  
usually described as a Boltzmann distribution of pure states with sharp momenta.
It can, however, be describe also as a mixture of rather sharply 
localized states, minimal wave packets, centered at points in the 
vessel  with the extension $a=(2mkT)^{-1/2}$  (see appendix).  
For H${}_2$–-molecules at room temperature the sharpness of  
localization of these pure states is $a=3 \cdot 10^{-9}$ cm.

	Other examples of particular interest are provided by 
experiments on persistent entanglement. A pair of particles is created 
in an entangled state and then far separated so that one of them is 
received in the lab of Alice, the other in the lab of Bob and
subjected there to simultaneous measurements by Alice and Bob in which 
neither knows what the other is doing.
Entanglement means that the particles are created in a two--particle 
quantum state which cannot be decomposed into a convex combination 
of pairs of single particle states. Such entangled or “non-separable” 
states obviously exist. Any pure two--particle state, which is not a simple 
product,  is an example. 
The ensemble of all particles received by Bob may be described by an 
impure one--particle quantum state $\rho_B$.
Since the twin particles are correlated due to their common birth it
is not surprising that the probability for a particular measuring 
result of Bob depends on the result of Alice’s measurement on the
twin. However, entanglement is more than ordinary correlation.  
An experimentally decidable test distinguishing entanglement from 
ordinary correlation was first presented by John Bell \cite{BL} and then 
proved in a more general context by Clausner, Horne and Shimony \cite{CHSH}. 
The difference is most easily demonstrated if one focuses on a degree 
of freedom described by a two--dimensional Hilbert space such as the 
polarization of a photon or the spin orientation for spin 
$1/2$. For 
chosen orientation of the measuring apparatus one only has two 
possible results which we denote by $\alpha= \pm 1$. Suppose now that 
a particle is endowed with some hidden objective property 
$\lambda$ and the 
joint probability in the ensemble of pairs of particles is 
given by a distribution function $\rho(\lambda_1, \lambda_2)$ 
which describes ordinary 
correlation between $\lambda_1$ and $\lambda_2$. 
In the original argument by Bell 
the hidden variable $\lambda$ was assumed to be “classical” i.e. to 
determine the outcome $\alpha$ for each measurement 
${\bf a}$. But as Clausner et 
al.~\cite{CHSH} showed, it suffices to assume a statistical connection 
between $\lambda$ and ${\bf a}$, $\alpha$ expressed by a probability
$w(\lambda ; {\bf a}, \alpha)$ or the expectation value
$$
\langle \,  {\bf a}; \lambda \, \rangle =
w(\lambda; {\bf a}, +) - w(\lambda; {\bf a}, -) \, .
$$
We note that 
$w(\lambda; {\bf a}, +) + w(\lambda; {\bf a}, -) = 1$
because in the measurement ${\bf a}$, one of the alternatives 
$\pm 1$ must occur. 
The joint probability for ${\bf a}, \alpha$; ${\bf b}, \beta$ is then
\begin{equation}
W({\bf a}, \alpha; {\bf b}, \beta) =
\int \! d \lambda_1 \, d \lambda_2 \, \rho(\lambda_1,\lambda_2)
\, w(\lambda_1; {\bf a}, \alpha) \, w(\lambda_2; {\bf b}, \beta) \, .
\end{equation}
For the expectation value in the joint measurement
$$
\langle \, {\bf a}; {\bf b} \, \rangle \equiv
w({\bf a}, +; {\bf b}, +) - w({\bf a}, +; {\bf b}, -) -
w({\bf a}, -; {\bf b}, +) + w({\bf a}, -; {\bf b}, -)
$$
one obtains the representation
\begin{equation}
\langle \, {\bf a}; {\bf b} \, \rangle = 
\int \! d \lambda_1 \, d \lambda_2 \, \rho(\lambda_1,\lambda_2) \, 
\langle \,  {\bf a}, \lambda_1 \, \rangle
\langle \,  {\bf b}, \lambda_2 \, \rangle \, .
\end{equation}
{}From this, together with the positivity and normalization of the 
distribution function $\rho(\lambda_1, \lambda_2)$ one obtains 
inequalities between 
expectation values for combinations of measurements with different 
orientations of the apparatuses,
\begin{equation}
|\langle \, {\bf a}; {\bf b} \, \rangle + 
\langle \, {\bf a}; {\bf b}^\prime \, \rangle +
\langle \, {\bf a}^\prime; {\bf b} \, \rangle -
\langle \, {\bf a}^\prime; {\bf b}^\prime \, \rangle|
\leq 2
\end{equation}
The experimentally observed  violation of this inequality shows that 
the assumption of an ordinary correlation between assumed 
properties $\lambda_1$, $\lambda_2$ is not tenable. 
Instead one has the following situation. 
If Bob receives the full information from Alice about what she has 
done and found in her measurements then he can first form the ensemble 
of all the particles whose twin was tested by Alice with the
orientation  ${\bf a}$ 
of the apparatus. The statistics of this ensemble show no difference 
from that of the full ensemble received. It is thus also described by 
the state $\rho_B$. If, however, Bob divides this into two subensembles 
according to Alice’s measuring result $\alpha = \pm 1$ then these 
subensembles define two orthogonal pure states which depend on the 
orientation of Alice’s device. It must be stressed that this has 
nothing to do with any physical effect of Alice’s measurement on the 
particles received by Bob. Nor is it important how fast the
information is transmitted. Bob and Alice can get together leisurely 
after the experiments are finished to evaluate their records. They
only have to establish the correct pairing of the events, which can be 
found for example from the records of the arrival times. 
No witchcraft is involved. It shows, however, that the 
pure state of the particle has no objective significance. 
It does not describe a property of an individual particle 
but only the defining information about the subensemble in 
which the particle is filed. This implies an enhancement of 
Bohr’s tenet mentioned in the introduction. Not only 
can we ``not assign any conventional attribute to an atomic 
object'' but we cannot even assign any individual state 
to the particle. This impossibility is at the root of the 
arguments about the non-existence of objective properties of a 
quantum system, among which the theorem by Kochen and Specker is perhaps 
the most convincing \cite{KS}
It puts in question our traditional picture of the 
reality of ``atomic objects'' (particles). 

	Nicola Maxwell has coined the term ``Propensiton'' 
for such an object \cite{M}. It propagates according to a deterministic 
law such as a Schr\"odinger equation which is invariant under 
time reversal. But it does not represent a real phenomenon. It 
is the carrier of  propensity contributing to probability
assignments. 
But probability for what? We must define a probability space i. e. a 
set of mutually exclusive possibilities (events). This is not 
determined by the beam alone. The particles in it need partners to 
produce events. In the simplest experimental set up described above 
the partner is a molecule in one of the detectors. The alternatives 
concern the choice of the detector which gives a signal. The result 
is interpreted as a position measurement of the particle after passing
the area of manipulation. In the collision experiment with crossed
beams 
at high energy the encounter of particles from the two beams may 
lead to a variety of different primary events. They are not directly
observable but can be reconstructed from the registration of the many
secondary 
ionization processes. The definition of the probability space of 
interest demands the enumeration of the distinctive attributes of 
such individual primary events. These consist of a localization region
in space- time and a channel (types of outgoing particles). Taken
together 
they constitute a new state carrying propensity for future events; a 
new deal. The states $\rho_1$, $\rho_2$ 
will usually describe particles with 
rather well defined momenta. The S-matrix and the specification 
of these momenta yield the probabilities for the various 
possible channels and also (due to the locality of the 
interaction) an intrinsic limitation of the relative distances between
the outgoing particles. They originate within a small common 
region in space. There is, however, no mechanism for 
distinguishing any point in the macroscopically large overlap 
area as the collision center which appears to be clearly 
visible in each individual event. To bring the theoretically 
predicted probability space in agreement with a sharp 
localization of the collision center one has to appeal to 
decoherence. Various factors can be made responsible for that. 
There is the limitation in the choice of possible subsequent 
events due to the given environment e.g. the cloud chamber. 
See the discussion of effective coherence length in section 2. 
This is a contingent decoherence. There is also the impurity 
of the state prepared in the initial beams and the 
non-uniqueness of their decomposition. But there remains a gray
zone. Ultimately, to get from probability space and 
probabilities to the emergence of an individual event, 
we need the principle of random realization discussed in the next section.

\subsection*{B \ Coherent, reversible processes vs.\  incoherent irreversible events}

There is a wide area in which coherence is preserved 
throughout all processes prior to actual detection. 
It includes all interference experiments, among them 
the diffraction of X-rays, electrons, neutrons by crystals; 
it includes manipulations of polarization or spin orientation, 
beam splitting and recombination of beams, used in the 
entanglement experiments. Their reversibility is demonstrated by the
so called quantum eraser. It even includes experiments in which an
atomic beam is crossed by laser light forcing the atoms to oscillate
between the ground state and an excited state. After several 
such encounters an interference between parts of the atomic 
beam having undergone different histories can still be observed. 

	The common feature of all these examples is that the 
back reaction from the microscopic object on the 
interaction partner is negligible. The interaction partners 
may be regarded as external fields. The processes remain 
in the realm of virtuality prior to detection. At the other 
end of the line there are the processes of high inelasticity 
and energy transfer including all detection processes leading 
to real, irreversible events.

\section{The principle of random realization}

In the foregoing sections we have drawn the picture of two 
stages of evolution of physical phenomena: coherent, 
deterministic, reversible propagation of propensitons 
followed by individually unpredictable, irreversible events. 
The existence of this second stage, though instinctively 
used by most physicists, is usually ignored or attributed 
to the acts and perception of an observer, to be explained 
in a theory of measurement. This remains too vague if 
it aspires to provide a general explanation. Let us restrict 
attention to particle physics. Then measurement theory 
reduces to the theory of detection processes.  
What do we detect? The presence of a particle? Or the 
occurrence of a microscopic event? We must decide for the 
latter. The detector fulfills two functions. It offers a 
target for a collision process, a microevent which is 
almost always the ionization of some molecule in the detector. 
Secondly it gives the amplification to visible dimensions 
via a chain reaction. The step from the virtual world of 
propensitons to a real fact must lie somewhere between the 
microevent and its amplification to a detector signal. 
For simplicity we assume that a single ionization 
process of one molecule can be clearly separated 
from subsequent processes so that we may consider this 
process already as a real event. The amplification poses 
no additional problem for the interpretation. 
Its mechanism is well understood and if we have perfect 
sensitivity there is a one-to-one connection between 
microevent and detector signal. 

The discussion shows that the standard use of the 
term ``observable'' does not really correspond to the 
needs of collision theory in particle physics. We do not measure 
a ``property of a microscopic system'', characterized
by a spectral projector of a self adjoint operator. Rather we are 
interested in the detection of a microscopic event.
The first task is to characterize the mutually exclusive 
alternatives for such an event. As mentioned in the last 
section this consists of a channel and a localization 
region defining together a new state. Of course there 
is a projector on this state. It is however not an operator in the
product of two single particle Hilbert spaces, but in the 
Fock space of outgoing particles and its determination 
is the main part of the theoretical effort. 

	This illustrates the first reason for the 
need to transcend the standard language. There the observable 
is assigned to ``the system'' (propensitons) as counterpart 
of a classical property translated to quantum theory by the 
machinery of quantization. Though the enormous historical 
importance and heuristic fruitfulness of this method is out 
of question it cannot be maintained in the regime of high 
energy collision processes. This is overcome in Quantum 
Field Theory. Most clearly in the algebraic approach by 
the concept of local observables, associated to regions 
in space-time, simulating detectors. But this is not enough. 
The idea that it is the observer who causes the realization of 
an event is not tenable either. The observer may produce the 
conditions by constructing cyclotrons, storage rings and 
sources which together determine the states of the 
crossed beams. And he constructs detectors to watch 
the results. But he has no influence on the emergence 
of resulting events. In particular the primary event 
being not directly observed is not the response to 
the measurement of a local observable. The theory 
provides the description of possible alternatives 
(the probability space) and the probabilities 
for the different possibilities. But we are still 
left to explain the emergence of individual facts 
whose appearance is governed by a statistical 
law which is intrinsic i.e. not due to any 
ignorance of hidden variables. This implies 
that the step from the virtual realm of 
propensities to reality is governed by a 
principle of random realization. The step 
is neither determined by previous history 
nor is it entirely free. The principle says 
that the theoretically predicted pattern will 
be realized  by a sequence of many individual 
events unpredictable in the individual case.

We may recognize a slight similarity to Niels Bohr’s 
somewhat mystical principle of an ultimate complementarity: 
the complementarity between space- time and causality. 
If causality refers to the deterministic propagation of 
propensitons and space-time stands for one of the 
essential attributes of events, namely their localization, 
this characterizes the same bipartition.  There are, however, 
essential differences. First, we would like to understand the 
term ``causality'' in a more liberal sense distinguishing it 
from determinism. Every event is connected by causal ties to 
preceding events \cite{Ha}. These are the propensitons. 
They leave, however, some freedom and do not give a 
strict command. Secondly, we do not regard the 
bipartition as a complementarity which allows us 
the choice to focus on one or the other aspect at a 
time. We need them both in succession.  

\vskip10mm
\noindent
{\large \bf Appendix}
\vskip 3mm
\noindent
{\bf Non uniqueness of decomposition of a general state 
-- Effective coherence length}
\vskip3mm
Let us start from a pure state with almost sharp momentum with mean 
value $\bar p$, mean position $\bar x$ and momentum uncertainty 
$\gamma^{-1/2}$ described by the wave function in momentum space
(we omit normalization factors)
\begin{equation}
\psi_1(p)= e^{-\frac{\gamma}{2}(p-\bar p)^2+ip\bar x},
\end{equation}
or in $x$ space
\begin{equation}
\psi_1(x)= e^{-\frac{1}{2\gamma}(x-\bar x)^2+i\bar p\,x}\,. 
\end{equation}
We consider a mixture of such states corresponding to an ignorance of the
precise values $\bar p$ and $\bar x$ expressed by the weight function
$e^{-\frac{\beta}{2}(\bar p -\hat p)^2-\frac{1}{2\alpha}{\bar x}^2}. $
The statistical matrix in $x$ space is $
\bra x'|\rho_1 |x\ket  = 
\int e^{-K_x} \rmd \bar x \rmd\bar p $ with
\begin{equation}
K_x= \frac {\bar x^2}{2\alpha} + \frac{\beta (\bar p -\hat p)^2}{2}
+\frac{1}{2\gamma}\left[(x-\bar x)^2+(x'-\bar x)^2\right] 
- i\bar p (x-x')\,.  
\end{equation}
Integration over $\bar p$ gives $\bra x'|\rho |x\ket = \int
e^{-K_1}\rmd \bar x$ with
\begin{equation}
K_1 = \frac {\bar x^2}{2\alpha}+\frac{(x-x')^2}{2\beta'}+
\frac{1}{\gamma}\left(\bar x-\frac{x+x'}{2}\right)^2- i\hat p (x-x')\, 
\end{equation}
with
\begin{equation}
\frac{1}{\beta'}=\frac{1}{\beta}+\frac{1}{2\gamma}\,.
\end{equation}

The same statistical matrix is obtained if we start from a pure state 
given by the wave function in $x$ space
\begin{equation}
\psi_2(x)=e^{-\frac{(x-\bar x)^2}{\beta'}+i \hat p\,x}
\end{equation}
and consider the mixture given with a weight factor 
$\exp(-\frac{\bar x^2}{2\alpha'})$. It leads at first sight to the 
following expression for the statistical matrix
\begin{equation}
\bra x'|\rho_2 |x\ket  = \int e^{-K_2} \rmd \bar x\,
\end{equation} 
with
\begin{equation}
K_2=\frac{\bar x^2}{2\alpha'}+\frac{1}{2\beta'}(x-x')^2
+\frac{1}{2\beta'}\left(\bar x- \frac{x+x'}{2}\right)^2 -i\hat p\,(x-x')
\end{equation}
After integration over $\bar x$ we see that $\rho_1=\rho_2$ provided
\begin{equation}
2\alpha+\gamma= 2\alpha'+\frac{\beta'}{2}\,.
\end{equation}
If $\gamma \gg \beta$, in the first version we then have a very large 
coherent extension $\gamma^{1/2}$ of the pure components. In the second 
version the effective coherence length $\beta'^{1/2}$ is much smaller 
corresponding to the much larger subjective ignorance of the momentum.
 

\vspace*{6mm}
\noindent{\bf \Large Acknowledgments} \\[2mm]
I want to thank Heide Narnhofer for many discussions and encouragement 
during several years. I am greatly indebted to  Detlev Buchholz  and Erhard 
Seiler for essential criticism and vital assistance during the last stages of this
work.

\end{document}